\documentclass[aoas,preprint]{imsart}

\RequirePackage[OT1]{fontenc}
\RequirePackage{amsthm,amsmath}
\RequirePackage{natbib}
\RequirePackage[colorlinks,citecolor=blue,urlcolor=blue]{hyperref}

\usepackage{verbatim,color,amssymb,epsfig}
\usepackage{graphicx,float,rotating,longtable,amsmath,amssymb,epsfig,color,lscape,subfigure,url}
\usepackage[normalem]{ulem}
\usepackage[percent]{overpic}
\usepackage{enumitem}
\usepackage{algorithm,algorithmic}

\def\S{{\cal D}}

\def\boxit#1{\vbox{\hrule\hbox{\vrule\kern6pt
			\vbox{\kern6pt#1\kern6pt}\kern6pt\vrule}\hrule}}

\def\wh{\widehat}

\def\bse{\begin{eqnarray*}}
	\def\ese{\end{eqnarray*}}
\def\be{\begin{eqnarray}}
\def\ee{\end{eqnarray}}
\def\bq{\begin{equation}}
\def\eq{\end{equation}}
\def\bse{\begin{eqnarray*}}
	\def\ese{\end{eqnarray*}}

\def\wh{\widehat}

\def\b1e{{\mathbf e}}

\def\bx{{\mathbf x}}
\def\bX{{\mathbf X}}
\def\by{{\mathbf y}}

\def\bbeta{{\boldsymbol{\beta}}}

\def\btheta{{\boldsymbol{\theta}}}
\def\bTheta{{\boldsymbol{\Theta}}}

\def\b1e{{\mathbf e}}

\def\by{{\mathbf y}}

\def\bz{{\mathbf z}}

\def\bx{{\mathbf x}}
\def\bX{{\mathbf X}}

\def\bZ{{\mathbf Z}}

\def\bone{{\mathbf 1}}

\def\bepsilon{{\boldsymbol \epsilon}}

\def\sqrtpl{\sqrt{p_{\ell}}} 
\def\pl{p_{\ell}} 
\def\sqrtpg{\frac{\sqrt{p_g}}{\sqrt{1+K}}} 
\def\pg{p_g} 

\arxiv{arXiv:0000.0000}

\startlocaldefs
\numberwithin{equation}{section}
\theoremstyle{plain}

\endlocaldefs

\begin{document}

\begin{frontmatter}
\title{
svReg: Structural Varying-coefficient regression to differentiate how regional brain atrophy affects motor impairment for Huntington disease severity groups
}
\runtitle{The Structural Varying-coefficient Regression}

\begin{aug}
\author{\fnms{Rakheon} \snm{Kim}\thanksref{m1}\ead[label=e1]{rkim@stat.tamu.edu}},
\author{\fnms{Samuel} \snm{Mueller}\thanksref{m2}\ead[label=e2]{samuel.mueller@sydney.edu.au}}
\and
\author{\fnms{Tanya P.} \snm{Garcia}\thanksref{m1}
\ead[label=e3]{tpgarcia@stat.tamu.edu}}


\affiliation{Texas A\&M University\thanksmark{m1} and University of Sydney\thanksmark{m2}}

\address{R. Kim and T. P. Garcia\\Department of Statistics\\
Texas A\&M University\\
College Station, TX, 77843-3143, U.S.A.\\
\printead{e1}\\
\phantom{E-mail:\ }\printead*{e3}
}

\address{S. Mueller\\School of Mathematics and Statistics\\
University of Sydney\\
Sydney, NSW 2006,
Australia\\
\printead{e2}}

\end{aug}

\begin{abstract}
For Huntington disease, identification of brain regions related to motor impairment can be useful for developing interventions to alleviate the motor symptom, the major symptom of the disease. However, the effects from the brain regions to motor impairment may vary for different groups of patients. Hence, our interest is not only to identify the brain regions but also to understand how their effects on motor impairment differ by patient groups. This can be cast as a model selection problem for a varying-coefficient regression. However, this is challenging when there is a pre-specified group structure among variables. We propose a novel variable selection method for a varying-coefficient regression with such structured variables. Our method is empirically shown to select relevant variables consistently. Also, our method screens irrelevant variables better than existing methods. Hence, our method leads to a model with higher sensitivity, lower false discovery rate and higher prediction accuracy than the existing methods. Finally, we found that the effects from the brain regions to motor impairment differ by disease severity of the patients. To the best of our knowledge, our study is the first to identify such interaction effects between the disease severity and brain regions, which indicates the need for customized intervention by disease severity.\\
\end{abstract}



\end{frontmatter}

\section{Introduction}
\label{sec:intro}

For Huntington disease, a genetically inherited neurodegenerative disorder, developing interventions to alleviate the symptoms 
of the disease is the goal of many clinical trials. One of the main symptoms of the disease is motor impairment 
\citep{biglan2009motor, paulsen2014prediction, reilmann2014diagnostic}
and the motor symptom is known to be related to regional brain atrophy, that is, the loss of cells in some brain regions \citep{aylward2013regional}. Hence, one interest in clinical trials is to identify which brain regions are associated with motor impairment and stop or slow atrophy of those regions to prevent motor impairment. For example, the clinical trial SIGNAL determines the effect of an antibody 
on the regional brain volumes and assesses the motor functions of the participants \citep{rodrigues2018huntington} by total motor scores (TMS), a score from 0 to 124 with higher indicating more severe impairment \citep{kieburtz2001unified}. 

\begin{figure}
	\scalebox{0.5}{
	\mbox{
		\subfigure{
			\begin{overpic}[width=4.9in,angle=0]
				{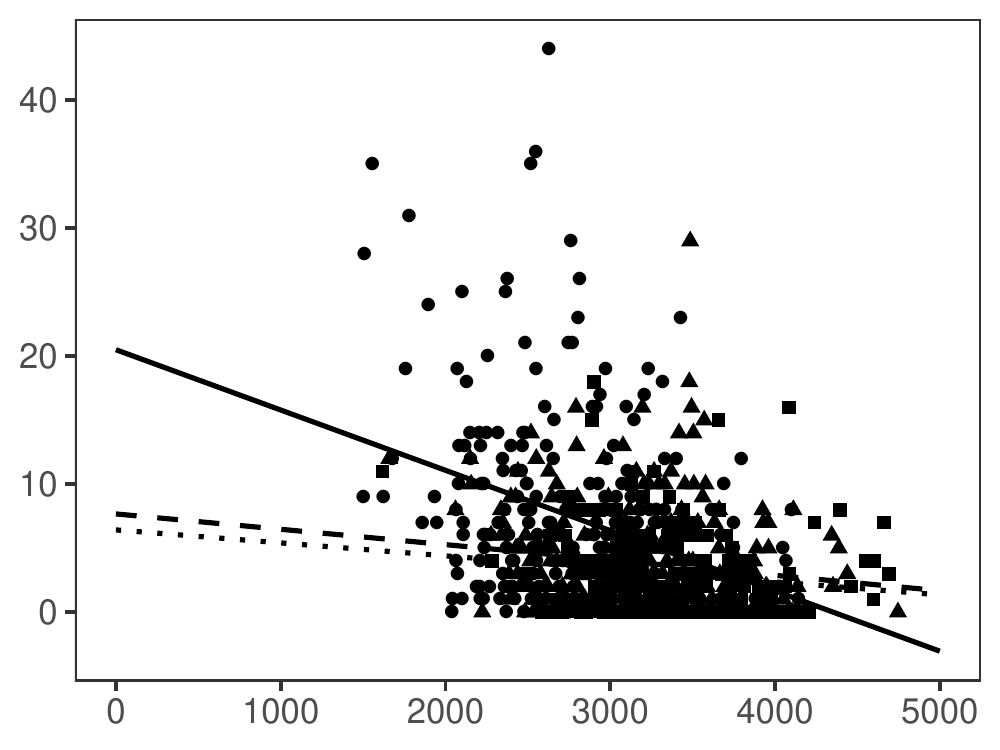}
				\put(-3,30){\rotatebox{90}{Total Motor Score}}
				\put(40,78){\LARGE{\uline{Left Caudate}}}
				\put(47,-2){Volume}
			\end{overpic}
		}
		\subfigure{
			\begin{overpic}[width=4.9in,angle=0]
				{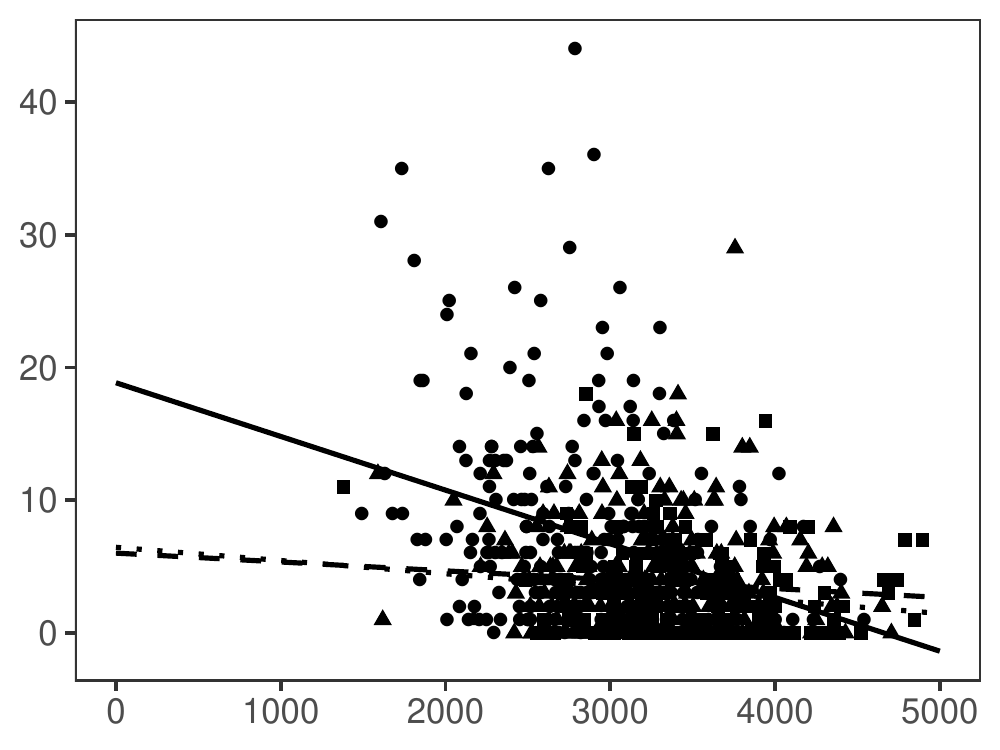}
				\put(40,78){\LARGE{\uline{Right Caudate}}}
				\put(47,-2){Volume}
			\end{overpic}
		}
		
	}
	}
	
	\vspace*{6mm}
	
	\scalebox{0.5}{
	\mbox{
		\subfigure{
			\begin{overpic}[width=4.9in,angle=0]
				{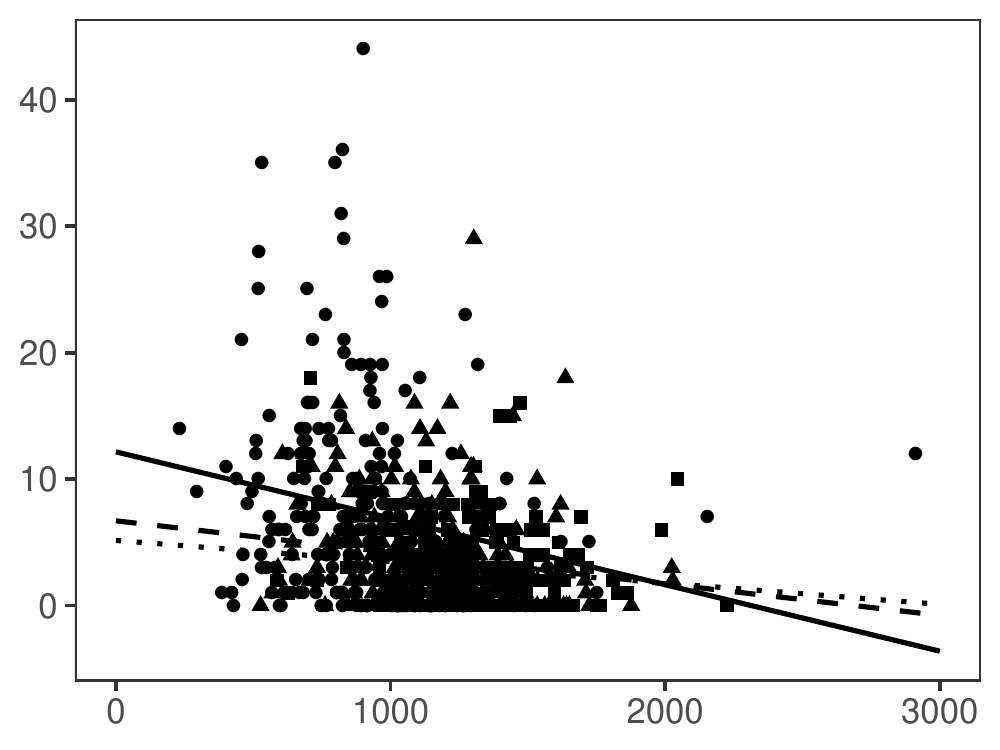}
				\put(-3,30){\rotatebox{90}{Total Motor Score}}
				\put(40,78){\LARGE{\uline{Left Pallidum}}}
				\put(47,-2){Volume}
			\end{overpic}
		}
		\subfigure{
			\begin{overpic}[width=4.9in,angle=0]
				{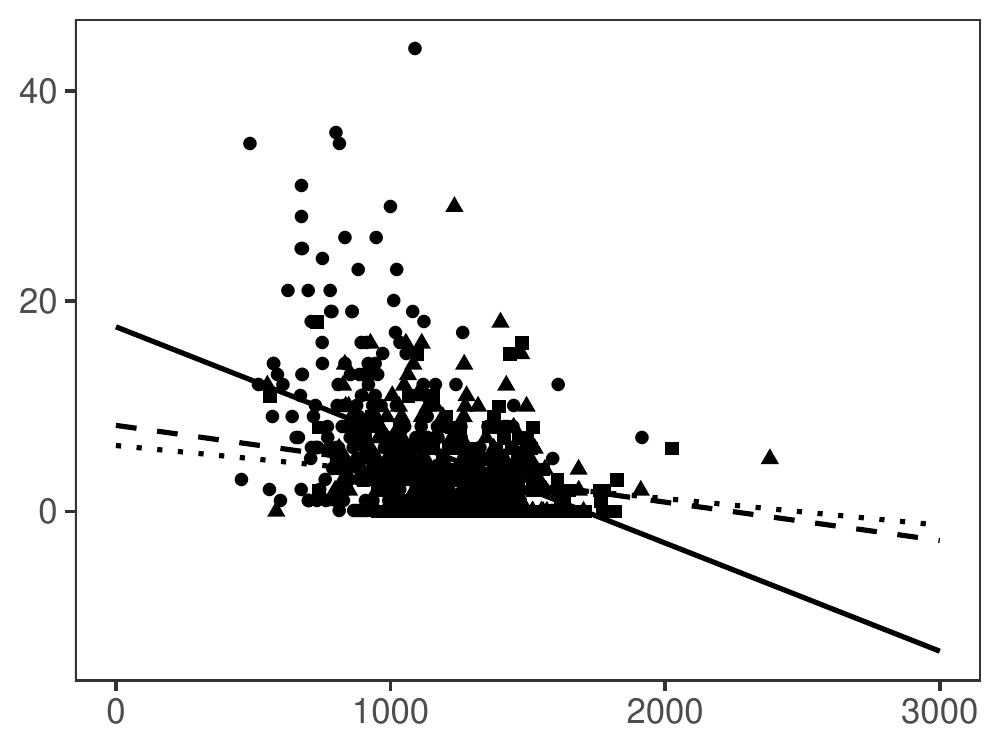}
				\put(40,78){\LARGE{\uline{Right Pallidum}}}
				\put(47,-2){Volume}
			\end{overpic}
		}
		
	}
	
	}
	
	\vspace*{6mm}
	
	\scalebox{0.5}{
	\mbox{
		\subfigure{
			\begin{overpic}[width=4.9in,angle=0]
				{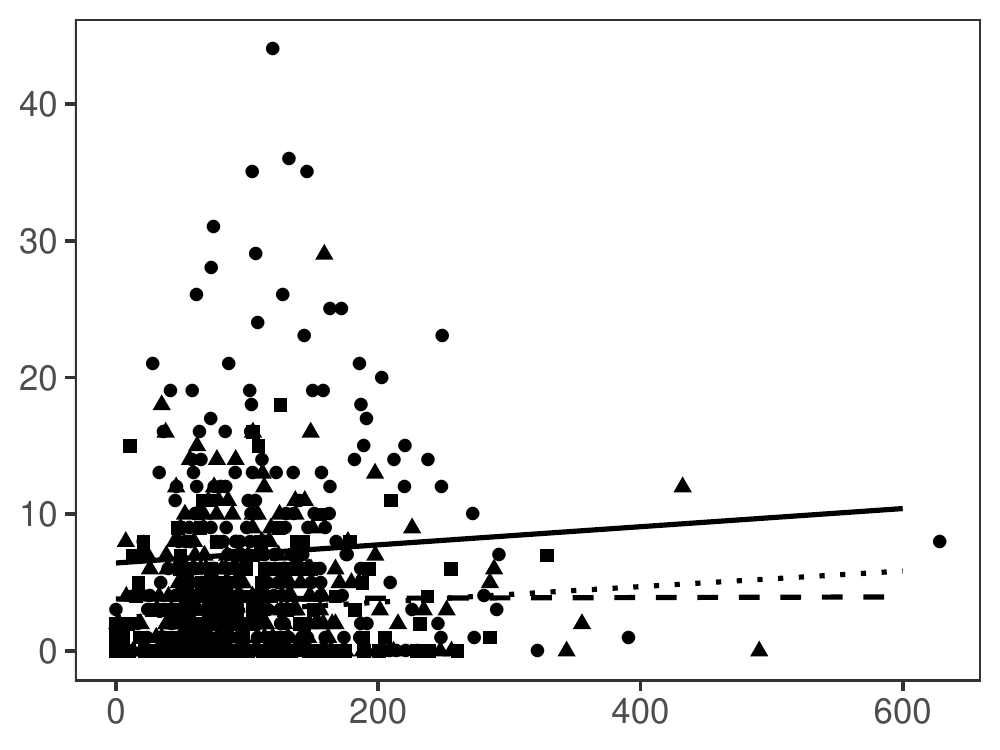}
				\put(-3,30){\rotatebox{90}{Total Motor Score}}
				\put(40,78){\LARGE{\uline{Left Vessel}}}
				\put(47,-2){Volume}
			\end{overpic}
		}
		\subfigure{
			\begin{overpic}[width=4.9in,angle=0]
				{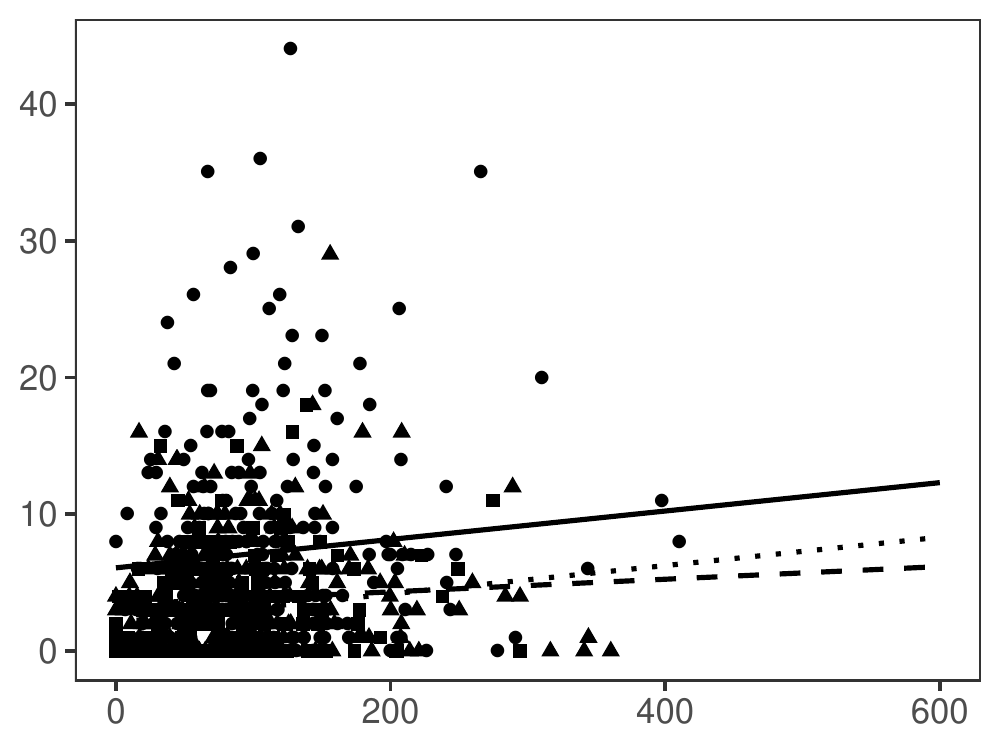}
				\put(40,78){\LARGE{\uline{Right Vessel}}}
				\put(47,-2){Volume}
			\end{overpic}
		}
		
	}
	
	}
	
	\vspace*{2mm}
	
	\scalebox{0.7}{
	\mbox{
	\begin{overpic}[width=4.5in,angle=0]
				{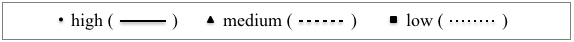}
			\end{overpic}
	}
	}

\vspace*{2mm}
\caption{Scatter plots between total motor score and volume of brain regions. Least squares fits by the group of scaled CAG-Age-Product (CAP) score, a measure of disease severity, are overlaid. Solid line is the least squares fit of the `high' disease severity group (circles), dashed line is the least squares fit of the `medium' disease severity group (triangles) and dotted line is the least squares fit of the `low' disease severity group (squares). Interaction effects between the volume of some brain regions (left caudate, right caudate, right pallidum) and CAP score are observed through different slopes of the least squares fit for each disease severity group. The difference in slopes is relatively small for the left pallidum and ignorable for the left and the right vessel. The correlation coefficient is 0.94 between the left caudate and the right caudate, 0.77 between the left pallidum and the right pallidum, and 0.48 between the left vessel and the right vessel.}
\label{fig:4.1}
\end{figure}
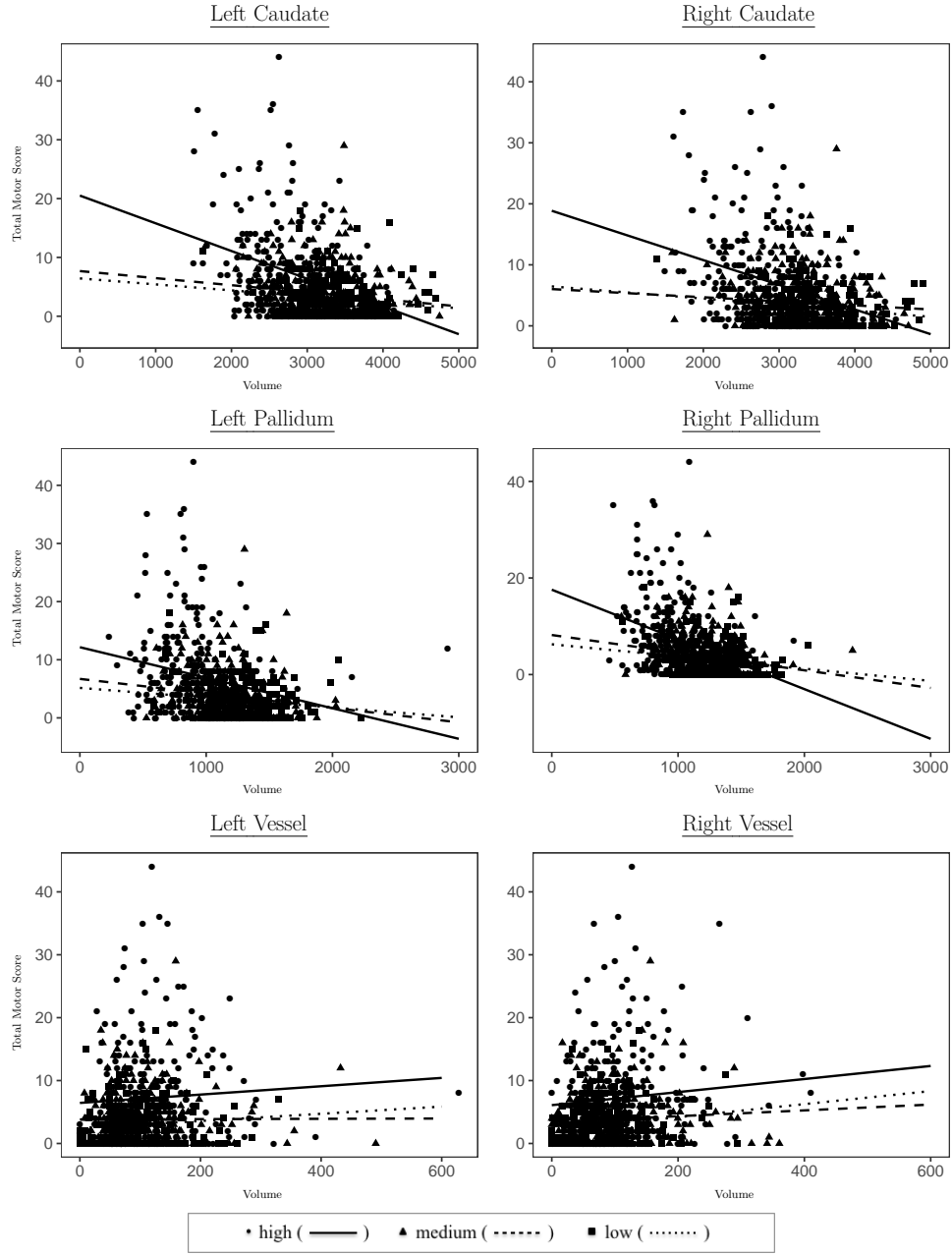

Although the relationship between the total motor score and the volume of brain regions is well understood \citep{aylward2013regional}, we observed that 
how the change of brain volumes affects the total motor score
may not be the same across all patients but vary for different groups of patients. For example, participants in a clinical trial can be categorized into three different groups (high/medium/low) by disease severity, a variable that indicates the risk of being diagnosed with Huntington disease in the next 5 years. In the top panels of Figure \ref{fig:4.1}, 
the effect from the reduction of caudate nucleus to the total motor score is larger for the high disease severity group than for other groups as observed by the steeper regression line.
This indicates that patients in the high disease severity group may need different interventions than patients in other groups since their motor function may deteriorate faster than others given a certain amount of change in caudate nucleus volume. 
Hence, in addition to the identification of brain regions related to motor impairment, understanding how their effects on motor impairment differ by patient groups will enable us to develop interventions customized for each patient group.

Statistically, identifying brain regions and understanding how their effects on motor impairment differ by patient groups can be cast as a model selection problem of a varying-coefficient model \citep{Hastie1993}. A varying-coefficient model is a regression model whose regression coefficients can vary by each individual or group of individuals. 
To be specific, consider a regression model with the total motor score as a response and the volume of brain regions as main predictors. In a varying-coefficient model, the regression coefficient of each brain region is not fixed but a function of other variables, called modifying variables. For example, if the disease severity is a modifying variable for a brain region, the regression coefficient of that region will take different value for each disease severity group so that we will end up with three different regression models, one for each group. Likewise, other demographic variables such as gender and years of education can also be considered as modifying variables which will divide the patients into smaller subgroups.

A varying-coefficient model is a special form of an interaction model where the interaction terms between main predictors and modifying variables are considered. In Figure \ref{fig:4.1}, the interaction effect between the volume of a brain region and the disease severity can be observed through difference in the slope of the regression line for each disease severity group. To the best of our knowledge, in the literature of Huntington disease, the disease severity and other demographic variables such as gender and years of education have been treated as covariates or control variables \citep{aylward2013regional, biglan2009motor, misiura2017cognitive}. However, their interaction effects with brain regions have not been investigated yet. In \citet{tabrizi2012potential} and \citet{paulsen2014clinical}, a different rate of change in brain regional volumes over time was observed for each disease severity group but the effect of the interaction on the total motor score was not considered.

Model selection of a varying-coefficient model includes two tasks: selection of main predictors and selection of modifying variables.
In the Huntington disease study, identifying brain regions related to motor impairment corresponds to the selection of main predictors. Understanding how the effects of those brain regions differ by patient groups corresponds to the selection of modifying variables where the possible candidates of modifying variables include disease severity, gender and years of education.

However, the literature on the varying-coefficient model has focused on variable selection of either the main predictors or the modifying variables, but rarely both. Among others, selection of main predictors has been explored when the modifying variable is a continuous variable \citep{Wang2008, Wei2011} or a categorical variable \citep{Gertheiss2012, Oelker2014}. In their work, only one modifying variable is considered so the interest is the selection of main predictors and whether each regression coefficient is fixed or not. 
Selection among multiple modifying variables has recently been explored through tree-based approaches \citep{Berger2017, Burgin2015, Wang2014}, which estimate a tree of modifying variables for each main predictor. However, these approaches focus on the selection of modifying variables and do not consider the selection of main predictors. \citet{tibshirani2019pliable} handles the variable selection of a varying-coefficient model by the \textit{pliable Lasso} (pLasso), a generalization of the Lasso (least absolute shrinkage and selection operator) that selects both the main predictors and modifying variables, simultaneously.

Additional consideration for Huntington disease application is that there are pre-specified group structures among main predictors and modifying variables. First, measurements of some brain regions can be grouped according to the structural information on a brain and they are often highly correlated. For example, the volume of the left caudate and the right caudate can be considered as a group. Due to their high correlation coefficient $(=0.94)$, as shown in the top panels of Figure \ref{fig:4.1}, the left caudate and the right caudate have similar negative relationship with the total motor score. Second, the disease severity is a categorical variable with three categories (low, medium and high), expressed in the design matrix for linear regression as a group of two binary dummy variables. Since each of these binary variables contains only partial information for one categorical variable, those two binary variables should be grouped. 

The pliable Lasso \citep{tibshirani2019pliable} is designed to work well when there is no pre-specified structure among the variables. However, if there is a group structure among the main predictors with high within-group correlation, we claim that the pliable Lasso may lead to inconsistent model selection by randomly selecting variables from those highly correlated variables as the usual Lasso suffers \citep{zhao2006model}. This problem of the pliable Lasso will be discussed with a simulation study in Section \ref{sec:simul}. Furthermore, modifying variables may also have a pre-specified group structure as appeared in our Huntington disease problem.
Since ignoring such group structure may lead to selecting more variables than necessary \citep{Yuan2006}, it is desirable to account for such group structure in model selection. 

In this paper, we propose the novel \textit{structural varying-coefficient regression} (svReg) for a varying-coefficient model with structured variables. This method imposes hierarchical group penalties on each group of main predictors and modifying variables to account for group structures among variables. Such hierarchical group penalties have been studied in other regression settings. To name a few, the group Lasso \citep{Yuan2006} and the sparse group Lasso \citep{Simon2013} address pre-defined group structure among regressors and the network Lasso \citep{Hallac2015} extends the group Lasso to a network setting. Some literature on structured variable selection \citep{Garcia2014, Garcia2013structured, Yuan2009} considers the structure between main effect terms and other variables such as interaction terms. However, simultaneous selection of main predictors and modifying variables for a varying-coefficient model with group-structured variables has not been explored yet.

Our svReg approach builds upon the pliable Lasso but differs significantly from that: first, a pre-specified group structure and the within-group correlation of the variables are considered in the svReg, whereas the pliable Lasso ignores such group structure; second, we discovered that weighting penalty terms differently leads to better variable selection performance by accounting for the different size of each group of main predictors and modifying variables. Hence, penalty terms are differently weighted in the svReg; third, when some modifying variables are selected in the model, the svReg algorithm identifies the groups of possibly significant modifying variables first and then selects variables from those identified groups to reduce false selection while the pliable Lasso selects variables from the set of all modifying variables. These important differences from the pliable Lasso allow the svReg to select relevant variables consistently and better screen irrelevant variables with higher prediction accuracy. This will be demonstrated in various simulation settings.


\section{The structural varying-coefficient regression model}
\label{sec:methods}

\subsection{Main Model}
We consider a varying-coefficient linear regression model with a response variable, $y$, and $p$ \textit{main predictors}, $\{x_j\}_{j=1}^{p}$, and $K$ \textit{modifying variables}, $\{z_k\}_{k=1}^{K}$, as below:
\begin{equation} \label{eq:2.1}
y = \sum_{j=0}^{p}   \left\{ \beta_j + f_j(z_{1},\ldots,z_{K})\right\} x_{j} + \epsilon,
\end{equation}
where $f_j(z_{1},\ldots,z_{K})$ is a function of modifying variables and $x_{0} = 1$, representing a potential intercept term and $\epsilon$ is the error term. In this model, $\{z_k\}_{k=1}^{K}$ modify how the $j$-th predictor $x_j$ affects the response $y$ through the function $f_j$. When $f(\cdot)\equiv 0$ for all $j=0,1,\ldots,p$, this reduces to a plain linear model with fixed coefficients. The inclusion of $f_j(z_{1},\ldots,z_{K})$ within the coefficient of $x_{j}$ allows the coefficient to vary depending on the modifying variables $z_{1},\ldots,z_{K}$. For independent subjects $i=1,\ldots,N$, we denote the response variable, $y_i$, and $p$ main predictors, $\{x_{ij}\}_{j=1}^{p}$, and $K$ modifying variables, $\{z_{ik}\}_{k=1}^{K}$.

In the Huntington disease study, our objective is to use the varying coefficient model to identify main predictors associated with total motor score ($y_i$) and understand how their effects on total motor score differ by patient groups where patients are grouped by modifying variables. Main predictors in our model will be selected from volume measures of 50 brain regions ($\{x_{ij}\}_{j=1}^p$, $p=50$). 
However, some of these regions are not independent because they are parts of a larger region. For example, caudate nucleus contains two parts, the left caudate and the right caudate, and correlation coefficient between their volumes is 0.94. That is, these measurements can be considered as a group of size two according to the structure of the brain. Likewise, lots of high correlations among the brain regions can be explained by the structural information of the brain.
Hence we consider the structure among the brain regions so that the 50 main predictors are grouped into 34 groups. 
Potential modifying variables will include gender ($z_{i1}$), years of education ($z_{i2}$) and disease severity ($z_{i3}, z_{i4}$), hence $K=4$. Here, disease severity is a categorical variable with 3 category levels (low, medium and high) depending on the likeliness of receiving a motor-diagnosis in the next five years. Hence, disease severity is expressed with two binary dummy variables, $z_{i3}$ and $z_{i4}$, and these two variables should be treated as grouped variables. Our proposed method will properly consider the group structure of the main predictors and the modifying variables by imposing group-wise penalty.

\subsection{Methodology}

We propose a novel modification to the \textit{pliable Lasso} \citep{tibshirani2019pliable} to account for 
potential structure among the variables (e.g., grouping between variables).
The pliable Lasso is a generalization of the Lasso for varying-coefficient models but it ignores potential structure among the variables, such as grouped main predictors (e.g., left and right caudate of the brain could be considered as one group) or grouped modifying variables (e.g., categorical disease severity group).
Ignoring such group structure and within-group correlation may lead to inconsistent model selection by randomly selecting variables from those highly correlated variables \citep{zhao2006model} or may lead to selecting more variables than necessary \citep{Yuan2006}. We thus propose a regression method with hierarchical penalties to account for grouped main predictors and grouped modifying variables.

Let $\by$ be the $N$ dimensional vector  $(y_1,\ldots,y_N)^T$ and let $\bX, \bZ$ be the $N \times p$ and $N \times K$ matrices containing main predictors and modifying variables respectively. Also, let $\bx_j$ be the $j$-th column of $\bX$, $\bz_k$ be the $k$-th column of $\bZ$ and let $\bone$ be a $N \times 1$ matrix of ones. We consider the following varying-coefficient linear model:
\begin{equation} \label{eq:2.2}
\by = \beta_0 \bone + \bZ \btheta_0 + \sum_{j=1}^{p} \left\{ (\beta_j \bone + \bZ \btheta_{j} ) \circ \bx_j \right\} + \bepsilon,
\end{equation}
where $\btheta_{j} = (\theta_{j1},\ldots,\theta_{jK})^T$. Here, $\circ$ is component-wise multiplication and captures the impact of the modifying variables by allowing coefficients to vary for each subject. In this model, the coefficient vectors $\{\btheta_{j}\}_{j=1}^{p}$ exist only within the coefficients of $\{\bx_j\}_{j=1}^{p}$. Hence, for $j=1,\ldots,p$, if $\bx_j$ turns out to be irrelevant (i.e. $\beta_j = 0$), we want $\btheta_j$ to be estimated as a zero vector. However, $\beta_j$ can take a nonzero value even if $\btheta_j$ is a zero vector, which results in a fixed coefficient for the $j$-th predictor. This feature of the varying-coefficient model raises the need to impose an ``asymmetric weak hierarchy" constraint: $\btheta_j$ can be nonzero only if $\beta_j$ is nonzero. 

Suppose the $p$ main predictors can be grouped into $L$ groups $(L \leq p)$ and the $K$ modifying variables can be grouped into $G$ groups $(G \leq K)$. Each group can contain one or more variables. In our Huntington disease application, there are 50 main predictors of brain regional volumes ($p=50$) and these predictors can be grouped into 34 groups of brain regions ($L=34$) according to the pre-specified structure of the brain. For the modifying variables, we have three groups of modifying variables ($G=3$): gender, years of education and disease severity. The first two groups contain one variable each. The disease severity group contains two dummy variables since disease severity is a categorical variable with three categories. Hence, there are four modifying variables ($K=4$).

We propose to optimize the following objective function:
\be \label{eq:2.4}
J^{*}(\beta_0, \btheta_0, \bbeta, \bTheta) = \frac{1}{2N} \sum_{i=1}^{N} r_i^2 + \lambda P_{\alpha}^{*} (\bbeta, \bTheta),
\ee
where $\bbeta=(\beta_1,\ldots,\beta_p)^T$, $\bTheta=(\theta_{jk})_{j=1,k=1}^{p,K}$ is a $p \times K$ matrix, and
\bse
r_i = y_{i}-\beta_0-\bz_{i \bullet}\btheta_0-\sum_{\ell=1}^{L} 
\bx_{i [\ell]} (\bbeta_{[\ell]} + \btheta_{[\ell] \bullet} \bz_{i \bullet}^T ),
\ese
where $\bz_{i \bullet}$ is the $i$-th row of $\bZ$, $\bx_{i [\ell]}$ is the $\ell$-th group of the main predictors for the $i$-th row of $\bX$, $\bbeta_{[\ell]}$ is a subset of $\bbeta$ for the $\ell$-th group of the main predictors, $\btheta_{[\ell] \bullet}$ is a subset of $\bTheta$ for the $\ell$-th group of the main predictors and
\begin{align*}
\lambda P_{\alpha}^{*} (\bbeta, \bTheta) &= (1-\alpha) \lambda \sum_{\ell=1}^{L} \sqrtpl{} \left\{||(\bbeta_{[\ell]}, \text{vec}(\btheta_{[\ell] \bullet}))||_{2} + \sum_{g=1}^{G} \sqrtpg{}||\text{vec}(\btheta_{[\ell] [g]})||_{2}\right\}\\ &+ \alpha \lambda \sum_{j,k} |\theta_{jk}|_{1},
\end{align*}
where $\pl{}$ is the size of the $\ell$-th group of the main predictors, $\pg{}$ is the size of the $g$-th group of the modifying variables, $\btheta_{[\ell] [g]}$ is a subset of $\bTheta$ for the $\ell$-th group of the main predictors and the $g$-th group of the modifying variables and $\text{vec}(\cdot)$ is a vectorization operator. Note that $\bbeta_{[\ell]}$ is a $\pl{}$ dimensional column vector, $\btheta_{[\ell] \bullet}$ is a $\pl{} \times K$ matrix and $\btheta_{[\ell] [g]}$ is a $\pl{} \times \pg{}$ matrix. This is similar to the pliable Lasso \citep{tibshirani2019pliable} but differs in three ways. First, if there is a pre-specified group structure among the main predictors, they can be grouped together so that they are selected or screened together in the variable selection procedure. Second, the penalty terms $||\text{vec}(\btheta_{[\ell] [g]})||_{2}$ in $\lambda P_{\alpha}^{*} (\bbeta, \bTheta)$ uses the $L_2$ penalty for each group of modifying variables rather than the group of all modifying variables. Lastly, the penalty terms are weighted differently depending on the size of the group of main predictors and modifying variables. We call this method the structural varying-coefficient regression (svReg). By having both $||(\bbeta_{[\ell]}, \text{vec}(\btheta_{[\ell] \bullet}))||_{2}$ and $||\text{vec}(\btheta_{[\ell] [g]})||_{2}$ in the penalty, the svReg meets the requirement of imposing asymmetric weak hierarchy for estimating a varying-coefficient model since it rules out the possibility that $\bbeta_{[\ell]}=0$ and $\btheta_{[\ell] [g]} \neq 0$ for any $g \in \{1,\ldots,G\}$. Also, it considers the group structure among the variables by using the $L_2$ penalty. The last term of the penalty gives sparsity to the individual coefficients $\theta_{jk}$'s.

\subsection{Optimization}

We use a blockwise coordinate descent to obtain the global minimum of equation (\ref{eq:2.4}). 
Denote $\bz_{i [g]}$ as a subset of $\bz_{i\bullet}$ for the $g$-th group of the modifying variables, $r_i^{(-\ell)} = y_i - \sum_{h \neq \ell} \left\{ \bx_{i [h]} (\bbeta_{[h]} + \btheta_{[h] \bullet} \bz_{i\bullet}^T) \right\}$ as the partial residual for the $\ell$-th group of the main predictors and $r_i^{(-\ell)(-g)} = r_i^{(-\ell)} - \bx_{i [\ell]} \sum_{m \neq g} \btheta_{[\ell] [m]} \bz_{i [m]}^T $ as the partial residual for the $g$-th group of modifying variables. The procedure for estimating $\{\beta_j\}_{j=0}^{p}$ and $\{\theta_{jk}\}_{j=0, k=1}^{p, K}$ is given in Algorithm \ref{alg1}. 

In the step 2-(2)-(b)-(i) of the Algorithm \ref{alg1}, if the variables of the $\ell$-th group are uncorrelated with variance one, that is $\sum_{i=1}^{N} \bx^T_{i [\ell]} \bx_{i [\ell]}/N=I$, the closed form solution of $\hat{\bbeta}_{[\ell]}$ is available as below:
\bse
\hat{\bbeta}_{[\ell]}= \text{max $\left\{ 1 - \frac{(1-\alpha) \lambda \sqrtpl{}}{|| R_\ell ||_2} ,0 \right\}$} \cdot R_\ell
\ese
where $R_\ell = \sum_{i=1}^{N} \bx_{i [\ell]}^T r_i^{(-\ell)}/N$. Note that this takes the similar form with the solution of the group Lasso proposed by \citet{Yuan2006}. Also, when there is only one predictor, say $j$-th predictor, in the $\ell$-th group, this solution is equivalent to the pliable Lasso \citep{tibshirani2019pliable} as below:
\bse
\hat{\beta}_j= \left( \frac{N}{\sum_{i=1}^{N} x_{ij}^2} \right) S_{(1-\alpha) \lambda} \left(\frac{1}{N} \sum_{i=1}^{N} x_{ij} r_i^{(-j)} \right).
\ese

However, in our Huntington disease study, the main predictors with group structure have high within-group correlation and no closed form solution for $\hat{\bbeta}_{[\ell]}$ is available. For the group Lasso, \citet{friedman2010note} proposed that the solution for $\hat{\bbeta}_{[\ell]}$ can be found by sequential optimization of each parameter in $\bbeta_{[\ell]}$. This one-dimensional search over the parameters in $\bbeta_{[\ell]}$ uses \texttt{optimize} function in the R package, which finds the minimum or maximum of a univariate function using golden section search and successive parabolic interpolation. We adopt this approach to compute $\hat{\bbeta}_{[\ell]}$ in 2-(2)-(b)-(i) of the Algorithm \ref{alg1}.

\begin{algorithm}
	\caption{Algorithm for the structural varying-coefficient regression}\label{alg1}
	\vspace{0.3 cm}
	\begin{enumerate}[label=\arabic*.]
		\item Given the initial estimate of 
		$(\bbeta, \bTheta)$,
		compute $\hat{\beta}_0$ and $\hat{\btheta}_{0}$ from the regression of the residual on $\bZ$.
		\vspace{0.3 cm}
		\item Given $\lambda$, $\alpha$ and convergence tolerance $\epsilon$, repeat the following procedure until convergence: $|J^{*(old)}(\hat{\beta}_0, \hat{\btheta}_{0}, \wh{\bbeta}, \wh{\bTheta}) - J^{*(new)}(\hat{\beta}_0, \hat{\btheta}_{0},\wh{\bbeta}, \wh{\bTheta})| < \epsilon$ where $J^{*}(\hat{\beta}_0, \hat{\btheta}_{0},\bbeta, \bTheta)$ is defined in equation (\ref{eq:2.4}).
		
		\begin{enumerate}[label=(\arabic*)]
		    \item Compute $J^{*(old)}(\hat{\beta}_0, \hat{\btheta}_{0},\wh{\bbeta}, \wh{\bTheta})$ with the current estimate of $(\hat{\beta}_0, \hat{\btheta}_{0},\wh{\bbeta}, \wh{\bTheta})$.\vspace{0.3 cm}
		    \item For a cycle of $\ell=1,2,\ldots,L$:
		    \begin{enumerate}[label=(\alph*)]
			    \item Check $(\hat{\bbeta}_{[\ell]}, \hat{\btheta}_{[\ell] \bullet}) = 0$ by checking $(\hat{\bbeta}_{[\ell]}, \hat{\btheta}_{[\ell] [g]}) = 0$ for all $g=1,2,\ldots,G$ as below:
			    \begin{align*}
			    \bigg\| \frac{1}{N} \sum_{i=1}^{N} \bx_{i [\ell]}^T r_i^{(-\ell)} \bigg\|_2 & \leq \sqrtpl{} (1-\alpha) \lambda, \text{ and }\\
			    \bigg\|S_{\alpha \lambda} \left( \frac{1}{N} \sum_{i=1}^{N} \text{vec}(\bx_{i [\ell]}^{T} \bz_{i [g]}) r^{(-\ell)(-g)}_i \right) \bigg\|_2 & \leq \sqrtpl{} (1+ \sqrtpg{})(1-\alpha)\lambda,
			    \end{align*}
			    where $S_\lambda(x) = x(1-\lambda/|x|)_{+}$ denotes the soft-thresholding operator.\vspace{0.2 cm}
			    
			    If all conditions are satisfied, set $(\hat{\bbeta}_{[\ell]}, \hat{\btheta}_{[\ell] \bullet}) = 0$ and skip to (d).
			    \item If $(\hat{\bbeta}_{[\ell]}, \hat{\btheta}_{[\ell] \bullet}) \neq 0$, check $\hat{\btheta}_{[\ell] \bullet} = 0$ by checking $\hat{\btheta}_{[\ell] [g]} = 0$ for all $g=1,2,\ldots,G$ as below:
			    \begin{enumerate}[label=(\roman*)]
				    \item First, compute $\hat{\bbeta}_{[\ell]}$ by one dimensional optimization of each parameter in $\bbeta_{[\ell]}$ until convergence as described in Section 2.3.
				    \item Then, check $\hat{\btheta}_{[\ell] \bullet} = 0$ given $\hat{\bbeta}_{[\ell]}$ by checking $\hat{\btheta}_{[\ell] [g]} = 0$ for all $g=1,2,\ldots,G$ as below:
				    \bse
				    \bigg\|S_{\alpha \lambda} \left\{ \frac{1}{N} \sum_{i=1}^{N} \text{vec}(\bx_{i [\ell]}^T \bz_{i [g]}) (r^{(-\ell)(-g)}_i - \bx_{i [\ell]} \hat{\bbeta}_{[\ell]}) \right\} \bigg\|_2 < (1-\alpha) \lambda \frac{\sqrt{p_g p_\ell}}{\sqrt{1+K}}.
				    \ese
    			\end{enumerate}
			    If (ii) is satisfied for all $g=1,2,\ldots,G$, set $\bbeta_{[\ell]} = \hat{\bbeta}_{[\ell]}$ and $\hat{\btheta}_{[\ell] \bullet} = 0$ and skip to (d).
			    \item If $\hat{\bbeta}_{[\ell]} \neq 0$ and $\hat{\btheta}_{[\ell] \bullet} \neq 0$ (i.e. if there exists $g^*$ such that $\hat{\btheta}_{[\ell] [g^*]} \neq 0$):
			    \begin{enumerate}[label=(\roman*)]
			        \item Use gradient descent to find $(\hat{\bbeta}_{[\ell]}, \hat{\btheta}_{[\ell] [NZ]})$ where $\btheta_{[\ell] [NZ]}$ denotes the set of nonzero $\btheta_{[\ell] [g]}$'s 
			        \item With the updated $\hat{\bbeta}_{[\ell]}$, check the condition in 2-(2)-(b)-(ii) for all $g=1,2,\ldots,G$ again to confirm whether $\btheta_{[\ell] [NZ]}$ contains the same set of $\btheta_{[\ell] [g]}$'s.
			        \item If the composition of $\btheta_{[\ell] [NZ]}$ changed, repeat (i)-(iii) with the updated $\btheta_{[\ell] [NZ]}$.
			    \end{enumerate}
			    \item Compute $\hat{\beta}_0$ and $\hat{\btheta}_{0}$ from the regression of the current residual on $\bZ$.
		    \end{enumerate}
		    \vspace{0.3 cm}
		    \item Compute $J^{*(new)}(\hat{\beta}_0, \hat{\btheta}_{0},\wh{\bbeta}, \wh{\bTheta})$ with the current estimate of $(\hat{\beta}_0, \hat{\btheta}_{0},\wh{\bbeta}, \wh{\bTheta})$.
		\end{enumerate}
	\end{enumerate}
\end{algorithm}

\subsection{Comparison with the Pliable Lasso}

The pliable Lasso proposed in \citet{tibshirani2019pliable} optimizes the objective function as below:
\begin{equation} \label{eq:2.3}
J(\beta_0, \btheta_0, \bbeta, \bTheta) = \frac{1}{2N} \sum_{i=1}^{N} r_i^2 + \lambda P_{\alpha} (\bbeta, \bTheta),
\end{equation}
where $r_i = y_{i}-\beta_0-\bz_{i \bullet} \btheta_0 -\sum_{j=1}^{p} x_{ij} (\beta_j + \btheta_{j \bullet} \bz_{i \bullet}^T )$ and
\bse
\lambda P_{\alpha} (\bbeta, \bTheta) = (1-\alpha) \lambda \sum_{j=1}^{p} (||(\beta_j, \btheta_{j \bullet})||_{2} + ||\btheta_{j \bullet}||_{2}) + \alpha \lambda \sum_{j,k} |\theta_{jk}|_{1}.
\ese
In this function, the group structure among the main predictors is not considered. This may lead to incorrectly screening true relevant variables when the variables are grouped variables with high within-group correlation as the Lasso which tends to randomly select variables among highly correlated variables \citep{zhao2006model}. Our proposed remedy for this inconsistent variable selection is to group the variables using the information on the group structure of the main predictors so that the grouped variables are selected into the model or screened from the model together. Also, all the $L_2$ penalty terms are weighted differently by $\sqrtpl{}$ in equation (\ref{eq:2.4}), accounting for different size of each group of the main predictors. This weight is analogous to the weight used in the group Lasso penalty \citep{Yuan2006}.

The penalty term $||\btheta_{j \bullet}||_{2}$ in equation (\ref{eq:2.3}) is for penalizing the group of all modifying variables as a whole and the term $|\theta_{jk}|_{1}$ is for penalizing each modifying variable. Hence, there is no consideration of the group structure among the modifying variables in equation (\ref{eq:2.3}). This may lead to spurious selection of irrelevant modifying variables as shown in our simulation study in Section \ref{sec:simul}. Assuming $L=p$ for simplicity, The svReg in equation (\ref{eq:2.4}) corrects this limitation by replacing the penalty term $||\btheta_{j \bullet}||_{2}$ with the terms $\{||\btheta_{j [g]}||_{2}\}_{g=1}^{G}$, which penalize each group of modifying variables with weight $\sqrt{p_g}/\sqrt{1+K}$. This weight accounts for the size of each group of modifying variables, $p_g$, and also finds balance between $||(\beta_j, \btheta_{j \bullet})||_{2}$ ($K+1$ parameters) and $||\btheta_{j [g]}||_{2}$ ($p_g$ parameters). 

\section{Simulation Study}
\label{sec:simul}

\subsection{Simulation Design}

We compared our structural varying-coefficient regression proposed in Section \ref{sec:methods} with the Lasso \citep{Tibshirani1996} and the pliable Lasso \citep{tibshirani2019pliable} in some simulation settings. First, we considered the case when we have both continuous and categorical modifying variables. Second, we additionally considered the correlation between main predictors so that the highly correlated main predictors can be considered as grouped variables.

\textbf{Setting 1 (Structured modifying variables):} We generated 50 standard Gaussian independent predictors with sample size $N=100$. We also generated twenty modifying variables: ten continuous variables, $z_{i1},\ldots,z_{i10}$, and ten categorical variables of three categories, $z_{i11},\ldots,z_{i30}$. Note that each categorical variable is expressed with two dummy variables, hence those two variables can be considered as grouped variables. The continuous modifying variables were generated from the standard Gaussian distribution. The categorical modifying variables were generated from the multinomial distribution with equal probability. The response was generated for $i = 1,\ldots,100$ from
\bse
y_i = x_{i1} + x_{i2} + (1 + z_{i1}) x_{i4} + (1 - z_{i2} + z_{i11} - z_{i12}) x_{i5} + \epsilon_i,
\ese
where $\epsilon_i \sim N(0,1)$.

\textbf{Setting 2 (Structured main predictors \& modifying variables):} As in Setting 1, we considered 50 main predictors in Setting 2. Let $\{X_i\}_{i=1}^{50}$ denote the $i$-th main predictor. We generated $X_3$ and $X_6$ to be correlated with $\{X_1, X_2\}$ and $\{X_4, X_5\}$, respectively, as follows:
\bse
x_{i3} = \frac{2}{3} x_{i1} + \frac{2}{3} x_{i2} + \frac{1}{3} \gamma_i \quad \text{and} \quad x_{i6} = \frac{2}{3} x_{i4} + \frac{2}{3} x_{i5} + \frac{1}{3} \delta_i
\ese
where $\gamma_i \sim N(0,1)$ and $\delta_i \sim N(0,1)$. Other main predictors were standard Gaussian with sample size $N=100$ and independent to each other. By this construction, $x_{i3}$ and $x_{i6}$ are normally distributed with mean 0 and variance 1 as other main predictors. Given the high correlation, we treated $\{X_1, X_2, X_3\}$ and $\{X_4, X_5, X_6\}$ as grouped variables when we fitted the svReg.
This simulation setting is similar to that used in \citet{zhao2006model} to create dependence between predictors in a model where the model selection result of the Lasso can be inconsistent. Modifying variables and the response were generated as in Setting 1.

We applied three methods to the simulated data: the Lasso, the pliable Lasso and the svReg. 
In the Lasso, all combinations of the interaction between main predictors and modifying variables are considered to avoid model misspecification since the true models contain interaction terms that need to be considered. Since both the Lasso and the pliable Lasso ignore the group structure of the main predictors and the modifying variables, the svReg is expected to perform better than those methods in selecting relevant main predictors and screening irrelevant categorical modifying variables.

We ran 100 simulations and used 10-fold cross-validation  in each simulation to find the optimal value of the tuning parameter $\lambda$. 
In the cross-validation, we used decreasing $\lambda$'s from 10 to 0.01 by 0.01 to find the solution path of the parameters. The $\lambda$ value which minimizes the mean squared error in the cross-validation was chosen for the model estimation. For the pliable Lasso and the svReg, the weight parameter $\alpha$ was fixed at 0.5. How to optimally choose $\lambda$ for the various methods considered is beyond the scope of this article and our choice here is one choice that allows a `fair' comparison of the three considered shrinkage procedures.

\subsection{Methods for Evaluation}

To evaluate the model selection performance of the three methods, we computed the false discovery rate (FDR)  \citep{Benjamini1995}, sensitivity and specificity, the average percentage of time variables are selected, and predictive accuracy as measured by the mean squared errors. We also visualise findings in so-called difference curves as introduced in \citet{Garcia2016}.

The FDR is defined as the ratio of the number of irrelevant variables selected over the total number of variables selected. 
It measures how likely the method makes ``false selection" so a high value of FDR is undesirable. Since the pliable Lasso ignores the group structure of the categorical modifying variables and treats the dummy variables separately, it is expected to select more irrelevant modifying variables spuriously than the structural varying-coefficient regression, leading to higher FDR. 

Sensitivity is a measure of the ``true positive rate" and it is the ratio of the number of relevant variables selected over the number of true relevant variables. Specificity is a measure of the ``true negative rate" and it is the ratio of the number of irrelevant variables screened over the number of true irrelevant variables. Both high sensitivity and high specificity are desirable. In addition, we report the geometric mean of sensitivity and specificity $(= \sqrt{\text{Sensitivity} \times \text{Specificity}})$ as used in \citet{Kubat1998}.

We also computed the average percentage of time the variables are selected. The average percentage is computed for the relevant variable group and irrelevant variable group of the main predictors and the modifying variables separately. High percentage of selection is desirable for the relevant variable groups and vice versa 
for the irrelevant variable groups. 

The predictive performance can be evaluated by the mean squared error (MSE) from the V-fold cross-validation. In V-fold cross-validation, the data is split into $(V-1)$ sets for a training set and a test set. The training set is used to fit a model (``training" step) and then, the fitted model is used for calculating the MSE of the response for the test set (``testing" step).

\subsection{Simulation Results}

Simulation results are reported in Table \ref{tb:3.1}. In this table, we compared the Lasso, the pliable Lasso and the structural varying-coefficient regression with respect to variable selection and prediction accuracy. All models were estimated with the tuning parameter $\lambda$ which gives the minimum MSE from 10-fold cross-validation. 

\begin{table}[ht]
\centering
\scalebox{0.82}{
\begin{tabular}{@{\extracolsep{0pt}} llllrrrrrr}
  \hline\\[-2.7ex]
 &  &  & &  \multicolumn{3}{c}{\uline{Setting 1}}  &  \multicolumn{3}{c}{\uline{Setting 2}}  \\
 &  &  & &  \multicolumn{3}{c}{(structured modifying}  &  \multicolumn{3}{c}{(structured main \&}  \\
 &  &  & &  \multicolumn{3}{c}{variables)}  &  \multicolumn{3}{c}{modifying variables)}  \\[1.3ex]
metric & covariates &  & & Lasso & pLasso & svReg & Lasso & pLasso & svReg\\ 
  \hline\\[-2.7ex]
Percentage & Main & Relevant & & 1.00 & 1.00 & 1.00 & 0.95 & 0.95 & 1.00 \\ [1.0ex] 
 of selection &  & Irrelevant & & 0.48 & 0.27 & 0.21 & 0.47 & 0.28 & 0.26\\ [1.0ex] 
   & Modifying & Relevant & continuous & 1.00 & 1.00 & 1.00 & 1.00 & 1.00 & 1.00\\ [1.0ex] 
   & & & categorical & 0.84 & 1.00 & 1.00 & 0.84 & 1.00 & 1.00\\ [1.0ex] 
   &  & Irrelevant & continuous & 0.72 & 0.78 & 0.57 & 0.64 & 0.80 & 0.68 \\ [1.0ex] 
   &  &  & categorical & 0.73 & 0.80 & 0.56 & 0.71 & 0.79 & 0.70 \\ [1.0ex] 
   \hline\\[-2.7ex]
\multicolumn{4}{l}{False discovery rate (FDR)} & 0.84 & 0.81 & 0.75 & 0.84 & 0.81 & 0.79 \\ 
   \hline\\[-2.7ex]
\multicolumn{4}{l}{Sensitivity} & 0.96 & 1.00 & 1.00 & 0.93 & 0.98 & 1.00 \\ 
\multicolumn{4}{l}{Specificity} & 0.43 & 0.54 & 0.66 & 0.45 & 0.53 & 0.59 \\ 
\multicolumn{4}{l}{Geometric mean of sensitivity and specificity} & 0.63 & 0.73 & 0.81 & 0.64 & 0.71 & 0.76 \\ 
   \hline\\[-2.7ex]
\multicolumn{4}{l}{Mean squared error (MSE)}  & 2.57 & 2.62 & 2.46 & 2.55 & 2.69 & 2.53 \\ 
   \hline\\[-2.7ex]
\end{tabular}
}
\caption{Simulation results for the Lasso, the pliable Lasso (pLasso) and the structural varying-coefficient regression (svReg). In Setting 1, 50 independent main predictors, 10 continuous modifying variables and 10 categorical modifying variables with 3 categories were generated. In setting 2, correlation between main predictors were additionally considered. All values are the average of the 100 simulations. MSE is computed with the tuning parameter $\lambda$ which gives minimum MSE from 10-fold cross validation. For the pliable Lasso and the structural varying-coefficient regression, $\alpha$ is set to 0.5.} 
\label{tb:3.1}
\end{table}

The pliable Lasso and the svReg select relevant modifying variables better than the Lasso since they correctly specify a varying-coefficient model and treat those modifying variables as the effect modifiers of the main predictors. Both methods also screen irrelevant variables better than the Lasso which leads to lower false discovery rate and higher specificity.

In Setting 1, the strength of the svReg over the pliable Lasso is observed in screening irrelevant variables, which in turn leads to lower FDR by up to 6\% points and higher specificity by up to 12\% points than the pliable Lasso. Hence, by considering the group structure among the modifying variables, the svReg identifies relevant variables correctly while making fewer inclusion of irrelevant variables than the pliable Lasso, which will eventually lead to a more parsimonious and correct model with easier interpretation.

In Setting 2, additional benefit of the svReg over the Lasso and the pliable Lasso can be found in consistent selection of relevant main predictors when those predictors are structured. As discussed in \citet{zhao2006model}, the Lasso fails to select the relevant main predictors consistently when the predictors are correlated and this is shown in Table \ref{tb:3.1} by the percentage of selection of the relevant main predictors $(=0.95)$ being less than one. Interestingly, similar pattern is observed in the pliable Lasso. Although model selection consistency of the pliable Lasso is not within the scope of this paper, this simulation result indicates that the pliable Lasso also suffers from the problem of inconsistent variable selection when the variables are highly correlated. On the other hand, in the svReg, those correlated variables were grouped to be selected or screened together. Hence, the svReg shows consistent result of variable selection for the relevant main predictors with 2\% point higher sensitivity than the pliable Lasso.

In terms of prediction accuracy, 
the cross-validation MSE of the structural varying-coefficient regression shows an improvement over the pliable Lasso by up to 6\% in both simulation settings. This reflects the gain from accounting for the group structure among the modifying variables. 

\begin{figure}
	\centering
	\scalebox{0.5}{

	\mbox{
		\subfigure{
			\begin{overpic}[width=5in,angle=0]
				{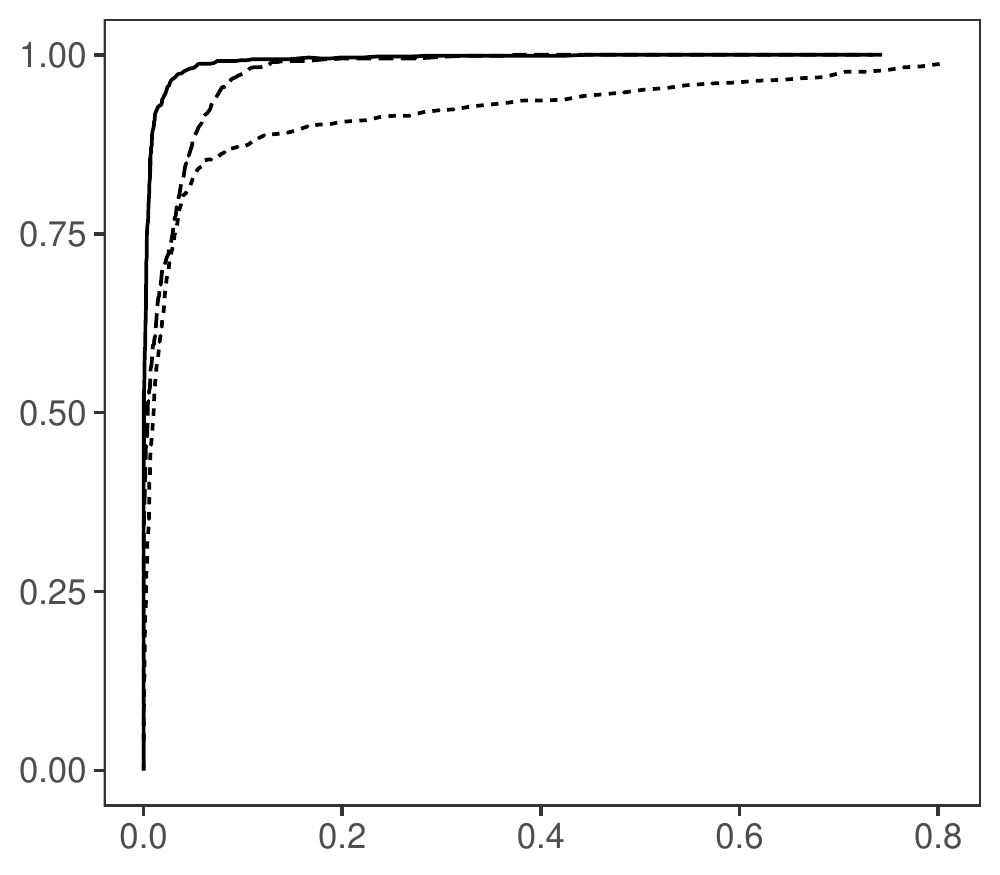}
				\put(-3,38){\rotatebox{90}{True Positive}}
				\put(46,-5){False Positive}
				\put(46,90){\LARGE{\uline{Setting 1}}}
			\end{overpic}
		}
		\subfigure{
			\begin{overpic}[width=5in,angle=0]
				{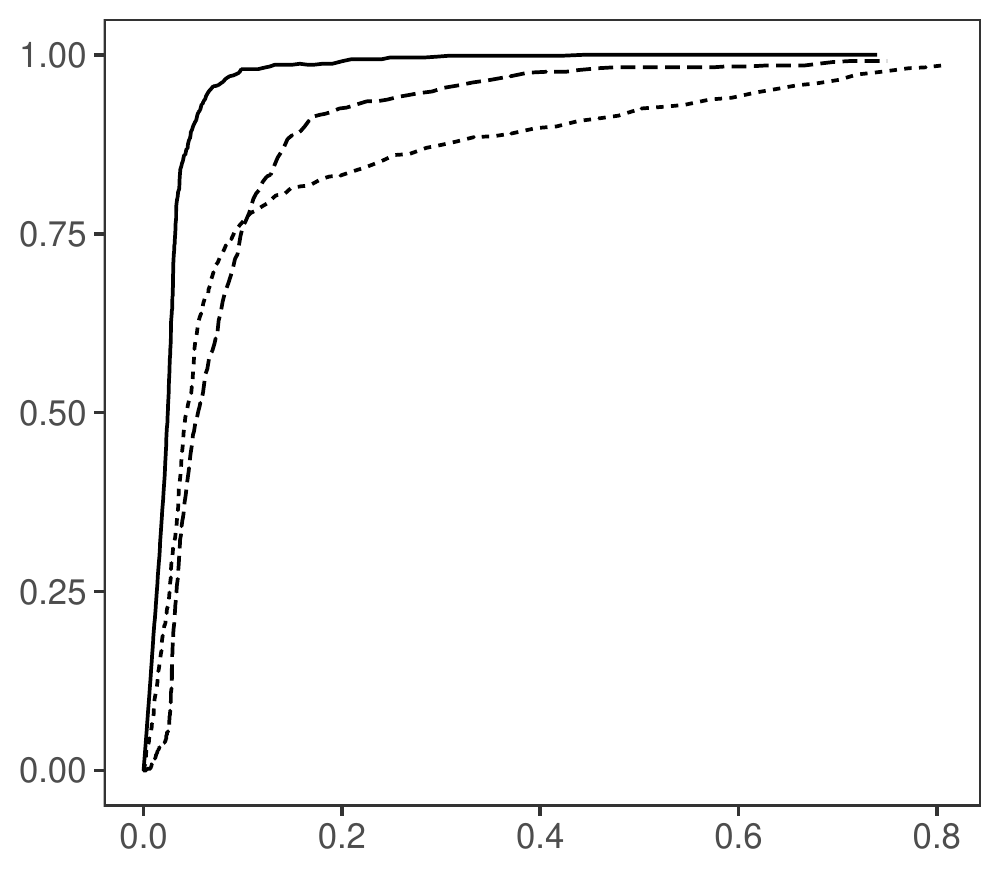}
				\put(46,-5){False Positive}
				\put(46,90){\LARGE{\uline{Setting 2}}}
			\end{overpic}
		}

	}
	}

\vspace*{6mm}
\caption{Receiver operating characteristic (ROC) curve of the Lasso (dotted curve), the pliable Lasso (dashed curve) and the structural varying-coefficient regression (solid curve) for Setting 1 and Setting 2. The structural varying-coefficient regression shows the lowest false-positive ratio for a fixed true-positive ratio. For the pliable Lasso and the structural varying-coefficient regression, $\alpha$ is set to 0.5.}
\label{fig:3.1}
\end{figure}
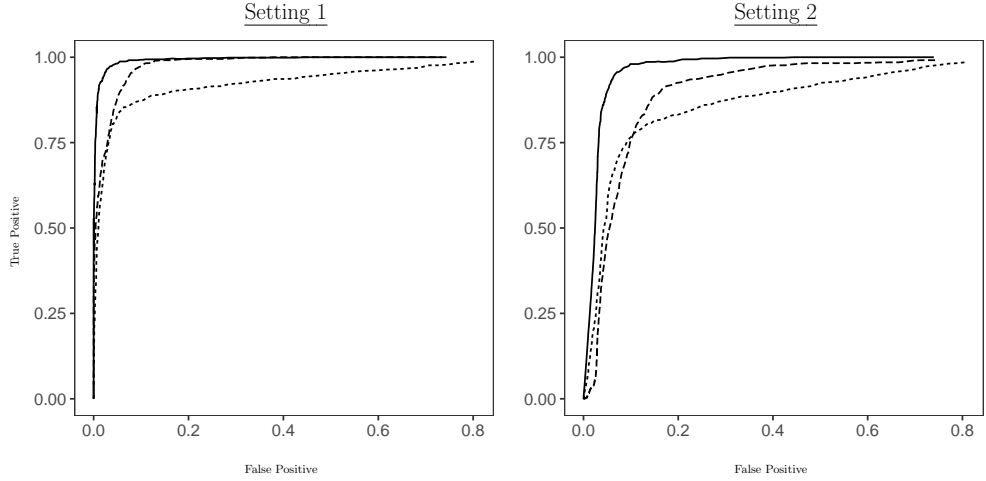

Figure \ref{fig:3.1} compares the receiver operating characteristic (ROC) curves of the Lasso, the pliable Lasso and the structural varying-coefficient regression. The ROC curve compares the true positive rate with the false positive rate over the different values of the penalty parameter, $\lambda$. True positive rate measures how well the method selects relevant variables and false positive rate measures the extent of incorrectly including irrelevant variables in the model. The structural varying-coefficient regression (solid red curve) shows higher true positive rate and lower false positive rate than other methods. Thus, we can conclude that structural varying-coefficient regression selects relevant variables more correctly while including fewer irrelevant variables than other methods.

\begin{figure}
	\centering
	
	\scalebox{0.5}{
	
	\mbox{
		\subfigure{
			\begin{overpic}[width=10in,angle=0]
				{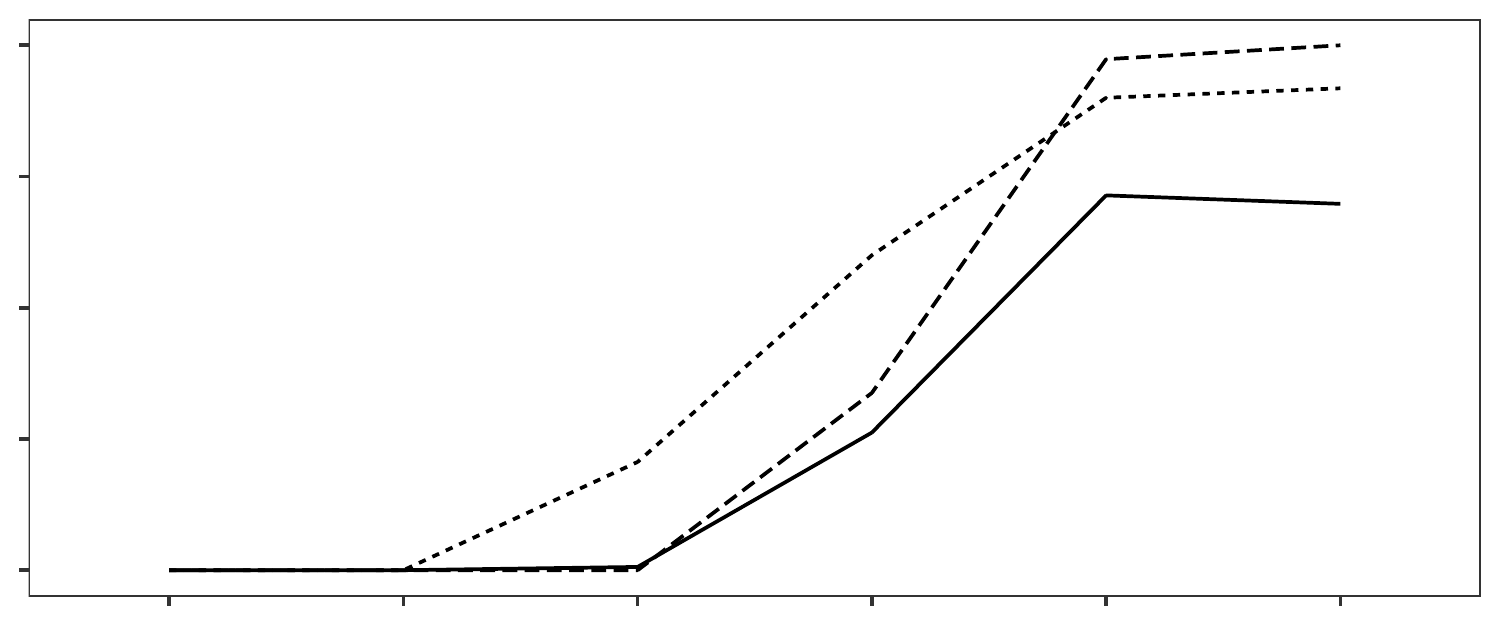}
				\put(-5,17){\large \rotatebox{90}{Difference\%}}
				\put(-2,36.8){80}
				\put(-2,28){60}
				\put(-2,20){40}
				\put(-2,11.5){20}
				\put(-1,3){0}
				\put(10,-1){main}
				\put(24,-1){continuous}
				\put(39,-1){categorical}
				\put(57,-1){main}
				\put(71,-1){continuous}
				\put(86,-1){categorical}
				\put(5,-5){\Large --------------------- relavant variables ---------------------}
				\put(53,-5){\Large -------------------- irrelavant variables --------------------}
				\put(47,42){\LARGE{\uline{Setting 1}}}
			\end{overpic}
		}

	}
	}
    
    \vspace*{20mm}
	
	\scalebox{0.5}{
	
	\mbox{
		\subfigure{
			\begin{overpic}[width=10in,angle=0]
				{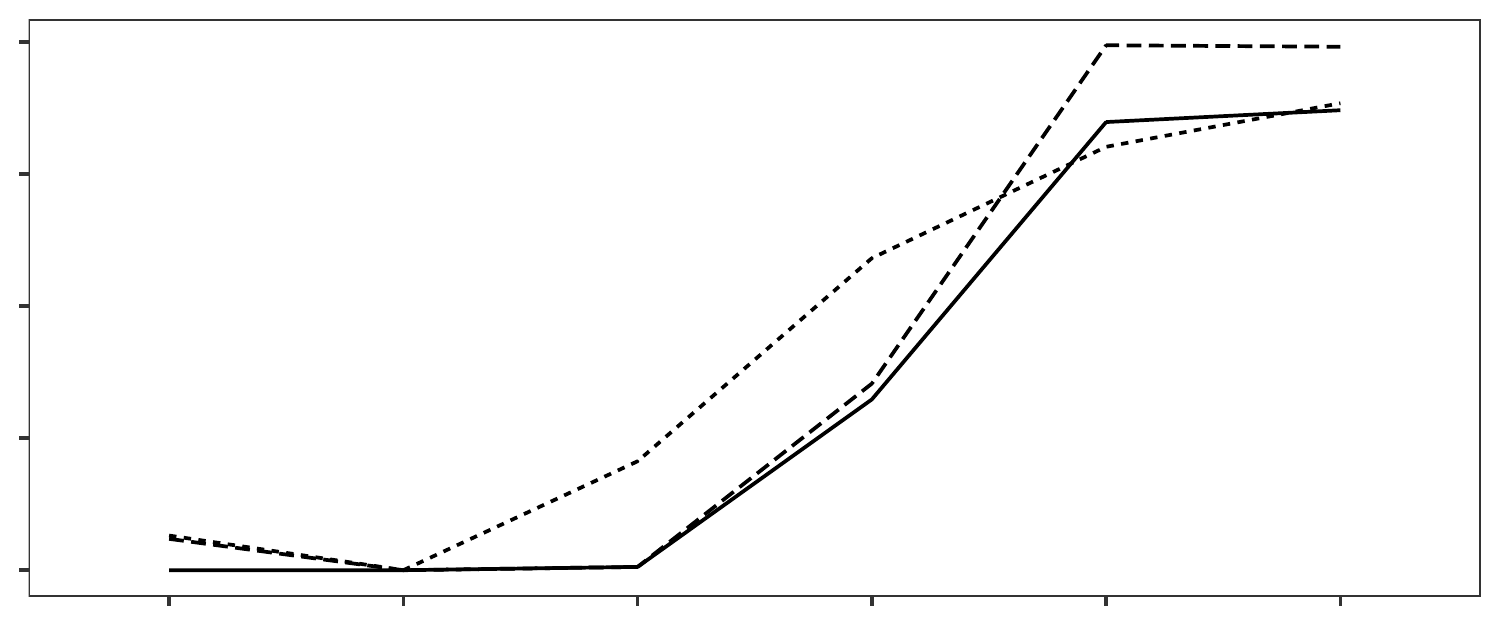}
				\put(-5,17){\large \rotatebox{90}{Difference\%}}
				\put(-2,33){75}
				\put(-2,23){50}
				\put(-2,13){25}
				\put(-1,3){0}
				\put(10,-1){main}
				\put(24,-1){continuous}
				\put(39,-1){categorical}
				\put(57,-1){main}
				\put(71,-1){continuous}
				\put(86,-1){categorical}
				\put(3,-5){\Large ------------------------ relavant variables -----------------------}
				\put(52,-5){\Large ----------------------- irrelavant variables -----------------------}
				\put(47,42){\LARGE{\uline{Setting 2}}}
			\end{overpic}
		}

	}
	}

\vspace*{10mm}
\caption{Difference curves of the Lasso (dotted curve), the pliable Lasso (dashed curve) and the structural varying-coefficient regression (solid curve) for Setting 1 and Setting 2. In a difference curve, a method with lower curve outperforms a method with upper curve in selecting relevant variables and screening irrelevant variables. In both settings, ``main" represents main predictors, ``continuous" represents continuous modifying variables and ``categorical" represents categorical modifying variables with 3 categories. The structural varying-coefficient regression generally shows lower difference than the Lasso and the pliable Lasso for both settings. For the pliable Lasso and the structural varying-coefficient regression, $\alpha$ is set to 0.5.}
\label{fig:3.2}
\end{figure}

Figure \ref{fig:3.2} compares the three methods by plotting the average percentage of selection using a difference curve, a visualisation introduced in \citet{Garcia2016}. In a difference curve, the average percentage of time selected for each group of variables is compared to the ``ideal" percentage of selection, which is 100\% for relevant variables and 0\% for irrelevant variables. That is, a better method in terms of variable selection has a lower curve in the plot. In both Setting 1 and Setting 2, the curve of the structural varying-coefficient regression is  below that of the pliable Lasso, which indicates that the svReg outperforms the pliable Lasso in selecting relevant variables and screening irrelevant variables. 

\subsection{Simulation without Structured Variables}

Although the motivation of developing the structural varying-coefficient regression was to deal with the structured main predictors and modifying variables, we can apply our method to the special case when there is no structure among variables. We compared the performance of the svReg with the pliable Lasso. For this purpose, 50 standard Gaussian independent main predictors and 20 binary modifying variables with equal probability were generated. The sample size $N$ was 100. The response was generated for $i = 1,\ldots,100$ from
\bse
y_i = x_{i1} + x_{i2} + (1 + z_{i1}) x_{i3} + (1 - z_{i2}) x_{i4} + \epsilon_i
\ese
where $\epsilon_i \sim N(0,1)$.

The result from this simulation is given in Table \ref{tb:3.2}. As in Table \ref{tb:3.1}, the svReg selects fewer irrelevant main predictors than the pliable Lasso by 5\% points and fewer irrelevant modifying variables by 17\% points. This leads to lower FDR and higher specificity for the structural varying-coefficient regression. Also, the prediction error of the svReg is lower than that of the pliable Lasso. 

\begin{table}[ht]
\centering
\scalebox{1.00}{
\begin{tabular}{@{\extracolsep{0pt}} lllrrr}
  \hline\\[-2.7ex]
metric & covariates &  & pLasso & svReg \\ 
  \hline\\[-2.7ex]
Percentage & Main & Relevant & 1.00 & 1.00 \\ [1.0ex] 
 of selection &  & Irrelevant & 0.21 & 0.16 \\ [1.0ex] 
   & Modifying & Relevant & 0.91 & 0.96 \\ [1.0ex] 
   &  & Irrelevant & 0.63 & 0.46  \\ [1.0ex] 
   \hline\\[-2.7ex]
\multicolumn{3}{l}{False discovery rate (FDR)} & 0.78 & 0.73 \\ 
   \hline\\[-2.7ex]
\multicolumn{3}{l}{Sensitivity} & 0.97 & 0.99 \\ 
\multicolumn{3}{l}{Specificity} & 0.67 & 0.75 \\ 
\multicolumn{3}{l}{Geometric mean of sensitivity and specificity} & 0.80 & 0.85 \\ 
   \hline\\[-2.7ex]
\multicolumn{3}{l}{Mean squared error (MSE)} & 1.59 & 1.54 \\ 
   \hline\\[-2.7ex]
\end{tabular}
}
\caption{Simulation results for the pliable Lasso (pLasso) and the structural varying-coefficient regression (svReg) when there is no structure among the main predictors or modifying variables. 50 independent main predictors and 20 continuous modifying variables were considered.  All values are the average of the 100 simulations. MSE is computed with the tuning parameter $\lambda$ which gives minimum MSE from 10-fold cross validation. For the pliable Lasso and the svReg, $\alpha$ is set to 0.5.} \label{tb:3.2}
\end{table}

The reason why the structural varying-coefficient regression outperforms the pliable Lasso for the variable selection purpose is related to the screening conditions for zero coefficients.
In the pliable Lasso, the screening condition for $(\hat{\beta}_j, \hat{\btheta}_{j \bullet}) = 0$ involves the calculation of the quantity as below:
\begin{equation} \label{eq:3.1}
\bigg\|S_{\alpha \lambda} \left( \frac{1}{N} \sum_{i=1}^{N} x_{ij} \bz_{i \bullet} r^{(-j)}_i \right) \bigg\|_2,
\end{equation}
and the screening condition is applied to the $L_2$-norm of the vector of coefficients for all modifying variables as a group (i.e., the size of this group is $K$). On the other hand, the corresponding condition for the structural varying-coefficient regression involves the calculation of the quantity as below:
\begin{equation} \label{eq:3.2}
\bigg\|S_{\alpha \lambda} \left( \frac{1}{N} \sum_{i=1}^{N} x_{ij} \bz_{i [g]} r^{(-j)(-g)}_i \right) \bigg\|_2.
\end{equation}
Note the $\sum_{i=1}^{N} x_{ij} \bz_{i [g]} r^{(-j)(-g)}_i$ takes a scalar value for a continuous modifying variable without any group structure with other modifying variables. In (\ref{eq:3.2}), each continuous modifying variable is treated as one group of variable (i.e., the size of each group is one) and the screening condition is applied to the coefficient for each modifying variable. Thus, the difference between (\ref{eq:3.1}) and (\ref{eq:3.2}) is that (\ref{eq:3.2}) will penalize each continuous modifying variable individually, while (\ref{eq:3.1}) will 
penalize all modifying variables as a group. Even if some elements of the coefficient vector are large and others are small, (\ref{eq:3.1}) can take large value which leads to possibly non-zero coefficients for all modifying variables whereas (\ref{eq:3.2}) will take small values for those elements.


Also, once some of the $\{\btheta_{j [g]}\}_{g=1}^G$ turn out to be zero, the svReg uses gradient descent procedure only for the nonzero $\btheta_{j [g]}$'s. This is not the case in the pliable Lasso where the gradient descent is performed for all $\{\theta_{jk}\}_{k=1}^K$ if $\btheta_{j \bullet}$ is nonzero. This allows the svReg to find the zero coefficients more efficiently than the pliable Lasso.

\section{Brain Regions Affecting Motor Impairment in Huntington Disease}
\label{sec:data_anal}

\subsection{Clinical Research Problem}
We applied our method to the Neurobiological Predictors of Huntington Disease (PREDICT-HD), a large observational study from 2001 to 2013 on potential neurobiological markers of Huntington Disease (HD). We focus on the data of $N=710$ subjects who are ``at risk" of HD with CAG (cytosine, adenine, guanine) repeats greater than or equal to 36. Subjects at risk means that they may or may not exhibit Huntington disease symptoms, whereas those with CAG repeat less than 36 is expected not to develop HD symptoms. The majority of the subjects were female (63.5\%). On average, the subjects were 40.5 years old, had 42.4 CAG repeats (ranges from 37 to 61), and had 14.2 years of education. 

In this study, participants enter the study at different phase of the disease. Hence, each participant is subject to different ``disease severity" or different proximity to HD diagnosis. As a measure of disease severity, we used the scaled CAG-Age-Product (CAP) score, the product of CAG repeats and age as proposed in \citet{Zhang2011}. CAP score is often used as a categorical variable to remove the within-group variability with three categories: low, medium and high. Participants categorized as ``high" are regarded as having high probability of being diagnosed with HD based on motor functions in the next 5 years. In our data, about 27\% of the subjects are categorized as ``low" with CAP score less than 0.67 and about 37\% of the subjects as ``high" with CAP score greater than 0.85.

In the PREDICT-HD study, the interest is to identify brain regions which are associated with motor impairment. As a measure of motor impairment, we used the total motor score (TMS), a measurement of the overall motor impairment ranging from 0 (no impairment) to 124 (high impairment). As covariates, we used the volume measures of brain regions. Also, as explained above, each subject has different disease severity. If we ignore this feature of the data, the effect from the brain regions on motor impairment will be mixed with the effect of the disease severity and the model will not capture the ``pure" effect of the brain regions. For this reason, CAP score has been used as another covariate or control variable \citep{Garcia2016, Zhang2011} in addition to the volume measures of brain regions. 

However, including the CAP score simply as another covariate assumes that the effects of brain regions on motor impairment are fixed regardless of the CAP score. This assumption is questionable since there may be a different pattern between, for example, the high CAP group and the low CAP group. In Figure \ref{fig:4.1}, the least squares regression line between total motor score and volume of brain regions were fitted for the high/medium/low CAP score groups separately. In the top left panel, covariate is the volume of the left caudate and the response variable is the total motor score. It can be clearly observed that the slope of the high CAP group (solid line) is different from that of low (dotted line) or medium (dashed line) CAP group. This difference in slope indicates that the effect of the left caudate on total motor score depends on whether a participant has high CAP score or not. 
On the other hand, in the bottom right panel where the covariate is the volume of the right vessel, the difference in slopes is not as clear as in the left caudate. These results indicate that the effects of some brain regions may differ by participant groups but other brain regions may not. 

Thus, our interest in this analysis is not only to identify brain regions associated with motor impairment but also to understand how their effects on motor impairment differ by participant groups. This can be achieved by fitting a varying-coefficient regression with the total motor score as a response, volumes of brain regions as main predictors and the CAP score as a modifying variable. In addition, we included gender and years of education as possible modifying variables since the effects of brain regions on motor impairment may also differ by participant groups defined by these variables. Since the CAP score data contains information of both age and CAG repeat by its definition, those two variables were not used as modifying variables.

For estimating the varying coefficient model, the pliable Lasso \citep{tibshirani2019pliable} and the svReg were used. As discussed in Section \ref{sec:methods}, the svReg can consider the pre-specified structure of the variables, whereas the pliable Lasso cannot. Since some main predictors represent the left part and the right part of a brain region (e.g. left caudate vs. right caudate), those main predictors were grouped in the svReg. Also, since the CAP score is expressed as a group of two binary dummy variables, those dummy variables were also regarded as grouped modifying variables in the svReg. Additionally, we considered the Lasso allowing for interaction terms to be selected as in Section \ref{sec:simul}. However, the Lasso is not appropriate for fitting a varying-coefficient model since some main predictors may not be selected even if their interaction terms are selected by the Lasso.
Hence, we compared the pliable Lasso and the svReg applied to the PREDICT-HD study. For both methods, the tuning parameter $\lambda$ was selected based on 10-fold cross-validation and the weight parameter $\alpha$ was set to 0.5.

\subsection{Analysis Results}

Table \ref{tb:4.1} summarizes results for the pliable Lasso (left table) and the svReg (right table). The first column for each method shows the fitted parameters, $\beta$, for the main predictors (brain regions) and the other columns show the fitted parameters, $\theta$, for the modifying variables (gender, years of education, CAP score) in the coefficient of each main predictor,
as defined in equation (\ref{eq:2.2}). Here, ``CAP(medium)" and ``CAP(high)" express the binary variable for the medium CAP score group and the high CAP score group, respectively.


From the nonzero $\theta$ estimates for basal ganglia (brain regions related to motor movements including caudate, putamen and pallidum), we can infer that the effects from these brain regions to motor impairment differ by CAP score groups. Particularly, the $\theta$ estimates for CAP(high) take negative values, meaning that high CAP score group has steeper slope as observed in Figure \ref{fig:4.1} than low or medium CAP score group. This indicates that the motor function of the high CAP score group may deteriorate faster than other groups given a certain amount of volume change in those brain regions.

Interestingly, the $\theta$ for CAP(high) in the coefficient of the left pallidum was determined to be zero by the svReg. This means that the effect of left pallidum on motor impairment may not differ significantly between the high CAP score group and other groups.
This is consistent with Figure \ref{fig:4.1} where the differences in slopes are relatively small for the left pallidum.
Note that, for the pliable Lasso, this $\theta$ estimate is zero simply because the main effect of the left pallidum was not selected. However, the main effect of the left pallidum may have been excluded randomly by the pliable Lasso due to its high correlation with the right pallidum. 
Hence, the pliable Lasso does not clearly tell us whether the effect of the left pallidum on motor impairment is the same across all participants or differ by disease severity groups whereas the svReg does.
The least squares regression of the total motor score on each brain region allowing for interaction with the disease severity also indicates that the interaction between each brain region and CAP(high) is significant for all regions in basal ganglia except for the left pallidum. These least squares regression results can be found in Table 1 of Supplement A.



The $\theta$'s for CAP(medium) in the coefficients of the putamen were determined to be zero by the svReg, whereas the pliable Lasso estimated a positive $\theta$ value for CAP(medium) in the coefficient of the right putamen. However, the $\theta$'s for CAP(medium) are expected to take negative values as those for CAP(high) because the baseline category is the low CAP group.
Thus, the positive $\theta$ parameter by the pliable Lasso may have been selected spuriously, meaning that the effect of right putamen on motor impairment may not differ significantly between the low CAP group and the medium CAP group. This can also be inferred from Figure \ref{fig:4.1} where the least squares fit slopes for the low group and the medium group were indistinguishable. Hence, the svReg resulted in selecting fewer irrelevant $\theta$'s than the pliable Lasso. This result is consistent with the simulation study in Section \ref{sec:simul} where the svReg selected fewer irrelevant variables than the pliable Lasso. Correct screening of irrelevant variables will not only result in models with smaller standard errors but also enable clinicians to avoid unnecessary segmentation of the patients in developing customized interventions for patient groups.

To the best of our knowledge, our study is the first to identify the interaction effect between CAP score and the volume of brain regions to motor impairment. This implies the genuine effect from the brain regions to motor impairment can be better understood when the CAP score is taken into account as a modifying variable in a varying-coefficient model. This knowledge can be useful in developing interventions or treatments which target specific group of patients. For example, a newly developed treatment may have some side effect. In this case, we may want to minimize the dosage of the treatment to reduce the risk of the side effect. From our research, we know that the high CAP score group will suffer more severe motor impairment than other groups given some change of the volume of caudate. If the degree of motor impairment is tolerable for low-medium CAP score group but not for high CAP score group, clinicians may need to use the treatment only for the high CAP score group or use different dosage for each group. 

\begin{table}
	\scalebox{1.0}{
	\mbox{
		\subfigure{
			\begin{overpic}[width=5.0in,angle=0]
				{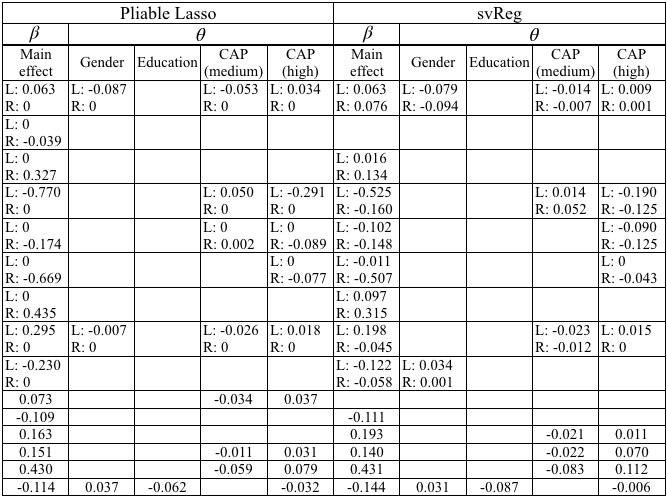}
				
				\put(-20,59){Lateral Ventricle}	\put(-22.5,54){Cerebellum Cortex}
				\put(-21,49){Thalamus Proper}
				\put(-11,44){Caudate}
				\put(-11.5,39){Putamen}	\put(-11.5,33.5){Pallidum}
				\put(-8,28){Vessel}
				\put(-18.5,23){Choroid Plexus}	\put(-25,18){CorticalWhiteMatter}
				\put(-16,14.2){3rd Ventricle}
				\put(-16,11.4){4th Ventricle}
				\put(-6.5,8.5){CSF}
				\put(-23,6){WM Hypointensity}
				\put(-17,3.5){Optic Chiasm}
				\put(-16,1){CC Posterior}
			\end{overpic}
		}
		
	}
	}

\vspace*{6mm}
\caption{Parameter estimates of the selected brain regions by the pliable Lasso (pLasso) and the structural varying-coefficient regression (svReg) for PREDICT-HD data. Parameter values are based on scaled data. Parameters not selected are shown as blank. The first column for each method contains the fixed part of the regression coefficients of main predictors ($\beta$'s in equation (\ref{eq:2.2})). The other columns represent the varying part of the regression coefficients of main predictors ($\theta$'s in equation (\ref{eq:2.2})). That is, the parameters from the second to fifth columns are the coefficients of the interaction terms between the brain regions (in row) and the modifying variables (in column). For the grouped brain regions (those with two lines), ``L" represents the left part of the corresponding brain region and ``R" represents the right part of the brain region. Tuning parameter $\lambda$ is selected from 10-fold cross-validation. $\alpha$ is set to 0.5.}
\label{tb:4.1}
\end{table}

\section{Discussion}
\label{sec:discuss}

In this paper, we proposed a new variable selection method for a varying-coefficient model with pre-specified group structure among variables. We showed in multiple simulation settings that ignoring this group structure among variables reduced the specificity by up to 12\% points and increased the false discovery rate by up to 6\% points. It also led to inconsistent selection of relevant main predictors when there is group structure with high within-group correlation and this lowered the sensitivity by 2\% points.
We applied our method to the Huntington disease study and found that the effect from basal ganglia to motor impairment differs by disease severity of the patients. Such knowledge suggests that different medical interventions might be needed depending on each patient's disease severity.

If other variables in addition to the disease severity are identified as relevant modifying variables 
in future study, that can be used for extending to the so called personalized interventions which account for the traits of each individual.
For example, if gender (male or female) and years of education (integer between 0 and 20) have turned out to be relevant modifying variables, the maximum number of possible models is 126 $(=3 \times 2 \times 21)$. Each of these models reflects the individual traits determined by the values of the three modifying variables for each patient and this individualized regression model will be useful for developing personalized interventions.

In our analysis, we considered only the linear combination of the modifying variables as the functional form of the varying coefficient ($f_j(\cdot)$ in equation (\ref{eq:2.1})). This is consistent with the basic setting discussed in \citet{tibshirani2019pliable} but both the pliable Lasso and the svReg can be generalized to consider other functional form of the varying-coefficient such as polynomials or splines. Particularly, considering the polynomials of modifying variables can be viewed as higher-order interaction model and can be implemented by adding higher-order modifying variable terms.

Our method is designed for a regression model. However, it can be extended to accommodate survival models or generalized linear models by changing the objective function in equation (\ref{eq:2.4}). For example, our method can be applied to Cox's proportional hazard model by adding the svReg penalty $\lambda P_{\alpha}^{*} (\bbeta, \bTheta)$ in equation (\ref{eq:2.4}) to the log partial likelihood of the hazard model. A similar attempt has recently been made by \citet{Du2018} for extending the pliable Lasso to the Cox's proportional hazard model. However, as with the pliable Lasso for a linear model, their method does not account for the pre-specified structure of the variables. The extension of the svReg to the hazard model is expected to select relevant variables consistently and screen irrelevant variables better than the method by \citet{Du2018} as was the case for the linear model settings and this will be future research.

\appendix

\section*{Acknowledgements}
This research is supported 
by
the National Institute of
Neurological Disorders and Stroke (NINDS; K01NS099343)  and Australian Research Council (DP170100654). This study used data from the PREDICT HD Study which received support from the National Institute of Neurological Disorders and Stroke  and collected by the PREDICT-HD investigators. 
We thank the 
PREDICT-HD investigators and respective coordinators who collected data and/or samples, 
as well as participants and their families who made this work possible.
The content is
solely the responsibility of the authors and does not 
represent the official views of
the National Institutes of Health.

\begin{supplement}
\sname{Supplement A}\label{suppA}
\stitle{Optimization details, additional analysis results and R package}
\slink[url]{https://github.com/rakheon/c2plasso}
\sdescription{The supplementary material contains the detailed derivation of the optimization criteria, regression analysis results for PREDICT-HD study, R code for simulation and R package.}
\end{supplement}

\bibliographystyle{imsart-nameyear}
\bibliography{biblio}

\end{document}